# The low-mass X-ray binary-millisecond radio pulsar birthrate problem revisited

DAI HaiLang[1,2] & LI XiangDong[2]

[1] Faculty of Science, Jiangsu University, Zhenjiang 212013, China

[2] Department of Astronomy, Nanjing University, Nanjing 210093, China

**We investigate the birthrate problem for low-mass X-ray binaries (LMXBs) and millisecond radio pulsars (MRPs) in this paper. We consider intermediate-mass and low-mss X-ray binaries (I/LMXBs) as the progenitors of MRPs, and calculate their evolutionary response to the cosmic star formation rate (SFR) both semi-analytically and numerically. With typical value (~1 Gyr) of the LMXB lifetime, one may expect comparable birthrates of LMXBs and MRPs, but the calculated number of LMXBs is an order of magnitude higher than observed in the Galaxy. Instead, we suggest that the birthrate problem could be solved if most MRPs have evolved from faint rather bright LMXBs. The former may have a population of ~ $10^4$ in the Galaxy.**

close binaries, general pulsars, star formation, X-ray stars

Millisecond radio pulsars (MRPs) have been suggested to be descendants of low-mass X-ray binaries (LMXBs) (Alpar et al.[1]). Formation of MRPs can be described by the "recycled" scenario (Bhattacharya & van den Heuvel[2]), in which mass accretion onto the neutron star (NS) from its binary companion spins it up to periods as short as a few milliseconds. This model has gained strong support in recent years from the discoveries of quasi-periodic kiloHertz oscillations and millisecond X-ray pulsations in a number of LMXBs (van der Klis[3]; Galloway[4], and references therein). However, statistical analysis have questioned whether the known LMXB population could produce the observed MRPs (e.g. Kulkarni & Narayan[5]; Cote & Pylyser[6]; Lorimer[7]). On one hand, the lifetime ($\tau_{\mathrm{MRP}}$ ~ 3-30 Gyr; Camilo et al.[8]) and total number ($N_{\mathrm{MRP}}$ ~ $4\times10^4$; Lorimer et al.[9]) of MRPs in the Galaxy require a birthrate of $N_{\mathrm{MRP}}/\tau_{\mathrm{MRP}}$ ~ $10^{-6}$ - $10^{-5}$ yr$^{-1}$. On the other hand, for the $N_{\mathrm{MRP}}$ ~ 100 LMXBs in the Galaxy with typical mass transfer timescale $\tau_{\mathrm{LMXB}}$ ~ 0.1-1 Gyr, the birthrate is estimated to be $N_{\mathrm{LMXB}}/\tau_{\mathrm{LMXB}}$ ~ $10^{-7}$ - $10^{-6}$ yr$^{-1}$, and this rate falls short of the MRP formation rate roughly by a factor of ~ 10 (e.g. Kulkarni & Narayan[5]; Cote & Pylyser[6]; Lorimer[7]).

White & Ghosh[10] and Ghosh & White[11] demonstrated that the above steady state arguments on the birthrate problem do not apply to evolving LMXB and MRP populations with a time-dependent global star formation rate (SFR) (Hopkins & Beacom[12], and references therein). By solving the differential equations describing the population of X-ray binaries, they found that an evolutionary scheme may account for the observed MRP/LMXB number ratio and birthrate ratio for the overall populations. However, favorable results can be obtained only when $\tau_{\mathrm{MRP}}$ ~ 0.1 Gyr is adopted. This value is smaller than typical ones inferred from both observations and theoretical works by roughly an order of magnitude.

There were suggestions that perhaps most of the current LMXBs have descended from systems with intermediate-mass ($\geq 1.5 M_\odot$) donor stars (Davies & Hansen[13]; King & Ritter[14]; Podsiadlowski & Rappaport[15-16]; Kolb et al.[17]; Tauris, van den Heuvel, & Savonije[18]). The aim of this paper is to check the influence of IMXB evolution on the formation of MRPs and the birthrate problem. In Section 1 we describe the model of White & Ghosh[10] and our modifications. Our numerically calculated results are presented in Section 2.

Finally, we make a brief discussion in Section 3.

## 1 Model

In the standard picture, LMXBs in the Galactic disk are produced by the evolution of primordial binaries containing a massive OB star and a low-mass star (see, e.g., Bhattacharya[19]). As described by White & Ghosh[10], the massive star in a primordial binary rapidly evolves to become a supernova (SN), resulting in an incipient NS in a post-supernova binary (PSNB). Subsequently, the PSNB evolves on a timescale $\tau_{PSNB}$ due to nuclear evolution or loss of orbital angular momentum, until the companion overflows its Roche lobe. The resultant LMXB evolves on a timescale $\tau_{LMXB}$ with mass transfer to produce a MRP. Finally, the MRP evolves on a timescale $\tau_{MRP}$. The evolutions of populations of PSNBs, LMXBs, and MRPs with a time-variable star formation rate $SFR(t)$ are given by the following equations,

$$\frac{\partial n_{PSNB}(t)}{\partial t} = \alpha_L SFR(t) - \frac{n_{PSNB}(t)}{\tau_{PSNB}} \quad (1)$$

$$\frac{\partial n_{LMXB}(t)}{\partial t} = \frac{n_{PSNB}(t)}{\tau_{PSNB}} - \frac{n_{LMXB}(t)}{\tau_{LMXB}} \quad (2)$$

$$\frac{\partial n_{MRP}(t)}{\partial t} = \frac{n_{LMXB}(t)}{\tau_{LMXB}} - \frac{n_{MRP}(t)}{\tau_{MPR}} \quad (3)$$

Here $n_{PSNB}$, $n_{LMXB}$ and $n_{MRP}$ are the number densities of PSNBs, LMXBs and MRPs in the Galaxy respectively, and $\alpha_L$ is a coefficient that determines the rate of formation of PSNBs per unit SFR (see White & Ghosh[10] for detail description). Solving Eqs. (1)-(3) one can obtain the evolution of these number densities with respect to cosmic redshift, if the form of $SFR(t)$ is given ($n_{PSNB}(0) = n_{LMXB}(0) = n_{MRP}(0) = 0$). This methodology, though suffers many simplified assumptions, can provide a straightforward picture of the evolution of various populations, and an estimation of the parameters on order of magnitude.

Since the formation probability of IMXBs with initial donor masses of 1.5-4$M_\odot$ is typically $\geq$ 5 times higher than that of standard LMXBs (Pfahl, Rappaport, & Podsiadlowski[20]), we assume that PSNBs with a low mass ($\leq 1.5\ M_\odot$) secondary evolve into standard LMXBs directly, and for the IMXB channel, the evolutions of populations of PSNBs, IMXBs, LMXBs, and MRPs in response to a time-dependent star formation rate $SFR(t)$ are given by

$$\frac{\partial n_{PSNB}(t)}{\partial t} = \alpha_I SFR(t) - \frac{n_{PSNB}(t)}{\tau_{PSNB}} \quad (4)$$

$$\frac{\partial n_{IMXB}(t)}{\partial t} = \frac{n_{PSNB}(t)}{\tau_{PSNB}} - \frac{n_{IMXB}(t)}{\tau_{IMXB}} \quad (5)$$

$$\frac{\partial n_{LMXB}(t)}{\partial t} = \frac{n_{IMXB}(t)}{\tau_{IMXB}} - \frac{n_{LMXB}(t)}{\tau_{LMXB}} \quad (6)$$

$$\frac{\partial n_{MRP}(t)}{\partial t} = \frac{n_{LMXB}(t)}{\tau_{LMXB}} - \frac{n_{MRP}(t)}{\tau_{MPR}} \quad (7)$$

Here, $n_{IMXB}(t)$ is the number density of IMXBs in the Galaxy and $\alpha_I$ is the formation rate of IMXBs. The ratio of $\alpha_I$ and $\alpha_L$ is taken to be 5 in this work. In Eqs. (1) and (4), $SFR(t)$ is described as the models of "peak" class with the form (Ghosh & White[11])

$$SFR(t) = 2[1 + \exp(\frac{z}{z_{max}})]^{-1}(1+z)^{p+(2z_{max})^{-1}} \quad (8)$$

where $z$ is the redshift, $z_{max} = 0.39$, and $p = 4.6$.

## 2 Results

Figures 1 and 2 show our calculated evolutions of LMXBs and MRPs with Eqs. (1)-(7): we have demonstrated the results in terms of the redshift $z$, which is in response to the global time $t$ by $t_9 = 13(z+1)^{-3/2}$, where $t_9$ is $t$ in units of $10^9$ yr, and a value of Hubble constant $H_0 = 50$ kms$^{-1}$ Mpc$^{-1}$ is adopted (Madau, Della Valle, & Panagia[21]). The dashed, dotted and solid lines denote the SFR, the number densities of LMXBs and MRPs, respectively. In each figure we list the calculated values of the number ratio $N_r = n_{MRP}/n_{LMXB}$ and the formation rate ratio $R_r = (N_{MRP}/\tau_{MRP})/(N_{LMXB}/\tau_{LMXB})$ of MRPs and LMXBs at $z = 0$. We take the values of $\tau_{PSNB} = 1$ and 0.1 Gyr for those evolving to LMXBs and IMXBs respectively, and $\tau_{IMXB} = 0.01$ Gyr, $\tau_{MRP} = 10$ Gyr. We find that changing these values within reasonable ranges does not influence the final output considerably. The most important parameter is the duration of LMXB phase, $\tau_{LMXB}$.

The results presented in Fig. 1 show that the time delay between LMXBs and the peak of SFR becomes considerably shorter due to the new formation channel from IMXBs, that is, LMXBs appeared earlier by ~1 Gyr in the cosmic evolution than expected by White & Ghosh[10]. We have $N_r$ ~ 20, $R_r$ ~ 2 when $\tau_{LMXB}$ ~ 1 Gyr (left panel), and $N_r$ ~ 250, $R_r$ ~ 2.5 when $\tau_{LMXB}$ ~ 0.1 Gyr (right panel). The latter seems to be compatible with the estimate of Lorimer[7] ($N_r$ ~ 400 and $R_r$ ~ 1) for the whole LMXB population. However, we note that $\tau_{LMXB}$ ~ 0.1 Gyr applies only for LMXBs with orbital periods of tens of days; for most observed LMXBs, $\tau_{LMXB}$ ~ 1 Gyr is likely to be more reasonable (Pfahl, Rappaport, & Podsiadlowski[20]). This will predict too many LMXBs in the Galaxy. It is then evident that the difference in the calculated and inferred birthrate ratios remains even with the inclusion of IMXB progenitors of MRPs. The reason may be that IMXBs spend most of their X-ray active lifetime as LMXBs, and consequently the duration of the X-ray active life is generally not much shorter than for true LMXBs.

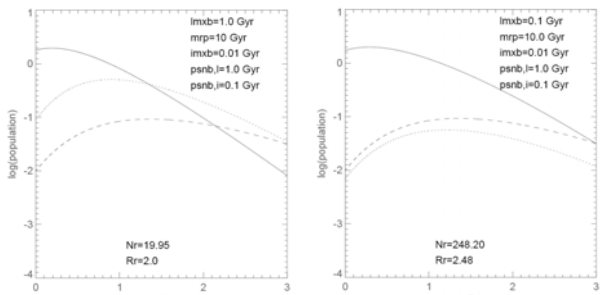

Fig. 1 Evolution of the number densities of LMXBs (dotted line) and MRPs (solid line) populations and the SFR (dashed line) as a function of the redshift z. The two panels demonstrate the results with various input timescales $\tau_{PSNB}$, $\tau_{IMXB}$, $\tau_{LMXB}$, and $\tau_{MRP}$, and the output values of the number ratio $N_r=n_{MRP}/n_{LMXB}$ and rate ratio $R_r=(N_{MRP}/\tau_{MRP})/(N_{LMXB}/\tau_{LMXB})$ at z=0.

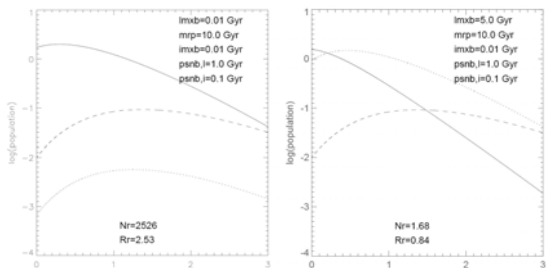

Fig. 2 Same as Fig. 1 but with the values of $\tau_{LMXB}$ changed.

To solve this problem, several authors have suggested that X-ray irradiation on the secondary in LMXBs can drive winds from the secondary (Ruderman et al.[22]) or cause the star to expand (Podsiadlowski[23]). These effects may lead to cyclic mass transfer, characterized by short episodes of enhanced mass transfer and long detached phases (e.g. Hameury et al.[24]), and significantly reduce the total X-ray lifetime. In the left panel of Fig. 2 we show the calculated results with $\tau_{LMXB}$ ~ $10^7$ yr. The value of $R_r$ does not change much from previous values, but $N_r$ increases by an order of magnitude, suggesting only ~10 LMXBs in the Galaxy.

The above arguments are based on the assumption that current MRPs have evolved from relatively bright LMXBs. However, it is possible that a significant fraction of LMXB population are transient, spending most of their lifecycle in quiescence (King[25]; van Paradijs[26]). Assuming transient LMXBs have a representative accretion rate of ~$10^{-10}$ $M_\odot$ yr$^{-1}$, corresponding to $\tau_{LMXB} \approx 5$ Gyr, we show the calculated evolutions in the right panel of Fig. 2. Now we have $R_r$ ~ $N_r$ ~ 1, implying that, if dim LMXBs are the main progenitors of the MRPs, there are probably ~ $10^4$ such objects in the Galaxy.

To clarify this point in more detail, we have calculated the formation of MRPs by using an evolutionary population synthesis code developed by Hurley et al.[27-28]. We take the initial mass function of Kroupa, Tout, & Gilmore[29] for the mass of the primary star (of mass $M_1$). For the secondary star (of mass $M_2$), we assume a uniform distribution of the mass ratio, $1 < q \equiv M_2/M_1 \leq 1$. A uniform distribution of ln$a$ is also taken for the binary separation $a$. During the core-collapse supernova explosions, we apply a Maxwellian distribution in the kick velocities with a mean of 265 kms$^{-1}$ imparted on the newborn NS (Hobbs et al.[30]). The common envelope parameter $\alpha$ is taken to be either 0.1 or 0.5. Finally, since a large fraction of the transferred mass during the previous I/LMXB phase may be lost from the systems rather than accreted by the NS (Bassa et al.[31], and references therein), we adopt an accretion efficiency parameter $\beta$ to denote the fraction of the mass lost by the donor star which is accreted by the NS. When mass transfer in an LMXB ceases, we assume that the system becomes a MRP with a white dwarf companion (when the initial mass of the secondary is large than $3M_\odot$, the reborn radio pulsar may be mildly recycled with a massive CO white dwarf companion). We stop the calculation when time *t* reaches 120 Gyr.

In Table 1 we list the calculated numbers of the LMXB and MRP populations for different values of $\alpha$ and $\beta$. Figure 3 shows the cumulative luminosity distributions of the total LMXBs (left) and of the LMXBs that have descended from IMXBs (right) in the case of $\alpha=0.1$ and $\beta=0.1$, which seems to be most compatible with both the observed luminosity distribution of LMXBs (Grimm et al.[32]) and the estimated number of MRPs in our Galaxy. From Table 1 and Fig. 3 one can clearly see that most LMXBs in our Galaxy are faint X-ray sources with X-ray luminosities $<10^{36}$ ergs$^{-1}$.

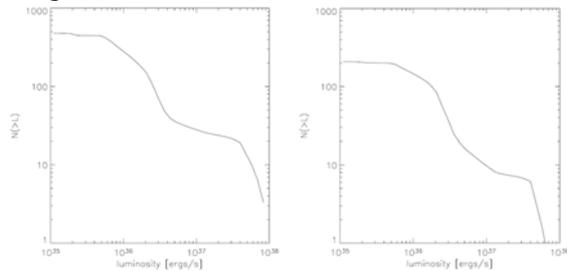

Fig. 3  Luminosity distributions of total LMXBs (left) and those evolved from IMXBs (right) based on our population synthesis calculations. The adopted parameters are: α=0.1, β =0.1.

Table 1: the predicted numbers of LMXBs and MRPs in the Galaxy.

| $N_{LMXBt}$[a] | $N_{LMXBi}$[b] | $N_{MRPt}$[c] | $N_{MRPi}$[d] | α | β |
|---|---|---|---|---|---|
| 9682 | 4678 | 51230 | 44288 | 0.1 | 0.1 |
| 13471 | 7895 | 74017 | 67075 | 0.1 | 0.5 |
| 22024 | 11347 | 179840 | 165085 | 0.5 | 0.1 |
| 40644 | 29769 | 206970 | 191730 | 0.5 | 0.5 |

[a]The number of total LMXBs
[b]The number of LMXBs descended from IMXBs
[c]The number of total MRPs
[d]The number of MRPs descended from IMXBs

## 3 Conclusion

Our work indicates that inclusion of IMXBs and their evolution with respective to the SFR do not immediately solve the LMXB-MRP birthrate problem. A critical parameter in the analysis is the mass transfer timescale in LMXBs. A typical value (~1 Gyr) predicts comparable birthrates of LMXBs and MRPs, but the number of LMXBs is an order of magnitude higher than observed in the Galaxy. Instead, we suggest that this problem could stem from incomplete sampling of LMXBs in the Galaxy - the majority of LMXBs are transient with low mass transfer rates (see Lorimer[7]; Pfahl, Rappaport, & Podsiadlowski[20]; Li[36]). Thus, MRPs may have evolved from dim LMXBs in the Galaxy, most of which have not been detected. This implies a population of ~$10^4$ LMXBs in the Galaxy.

If most MRPs really have evolved from transient LMXBs, the related question is whether sufficient mass has been accreted by the NS to ensure field decay and spin-up to milliseconds. Investigation by Xu & Li[37] on (sub)thermal mass transfer in IMXBs shows that in most cases an NS can accrete $\leq 0.1 M_\odot$ mass rapidly during this phase. This amount of mass is able to decrease the neutron star magnetic fields down to ~ $10^8$-$10^9$ G, and spin the star up to milliseconds. During the subsequent LMXB phase, the mass accretion is likely to be transient. In quiescence the accretion rates are so low that the NS may eject most of the accreting gas that reaches its magnetosphere and experiences a spin-down (propeller) torque; during outbursts the accretion rates onto the NS are orders of magnitude higher, the accreting gas is able to reach the NS surface and cause it to spin up again. For a sufficiently long mass transfer time the equilibrium spin periods are determined by the balance of the spin-up during outburst and spin-down during quiescence. For weak magnetic fields ($\leq 10^9$ G) the values of the equilibrium spin periods are shown to be in the range of 1-10 ms (Menou et al.[38]), consistent with observations of millisecond pulsations in LMXBs.

*This work was supported by 973 program 2009CB824800.*